\documentclass[11pt]{llncs}
\usepackage{graphicx}
\usepackage{cite}
\usepackage{url}
\usepackage{times}
\usepackage{mathptm}

\DeclareSymbolFont{AMSb}{U}{msb}{m}{n}
\DeclareSymbolFontAlphabet{\Bbb}{AMSb}

\def\Z{\ensuremath{\Bbb Z}}

\makeatletter
\def\hb@xt@{\hbox to }
\makeatother

\let\oldendproof\endproof
\def\endproof{\qed\oldendproof}

\def\Sc{\mathop{\rm Sc}}

\begin{document}

\title{The Lattice Dimension of a Graph} 

\author{David Eppstein}

\institute{Computer Science Department\\
School of Information \& Computer Science\\
University of California, Irvine\\
\email{eppstein@uci.edu}}

\maketitle   

\begin{abstract}
We describe a polynomial time algorithm for, given an undirected graph $G$, finding the minimum dimension $d$ such that $G$ may be isometrically embedded into the $d$-dimensional integer lattice~$\Z^d$.
\end{abstract}

\section{Introduction}

Geometric representations of graphs \cite{Lau-TR-94,LovVes-PEM-99} have been much studied for the insight they provide into the graph algorithms, graph structure, and graph visualization.
We consider here the following representation problem: for which unweighted undirected graphs can we assign integer coordinates in some $d$-dimensional space $\Z^d$, such that the distance between two vertices in the graph is equal to the $L_1$-distance between their coordinates?
We call the minimum possible dimension $d$ of such an embedding (if one exists) the
{\em lattice dimension} of the graph,  and we show that the lattice dimension of any lattice-embeddable graph may be found in polynomial time.

\section{Related Work}

Recently, Ovchinnikov~\cite{Ovc-04} showed that the lattice dimension of any tree is exactly $\lceil\ell/2\rceil$, where $\ell$ denotes the number of leaves of the tree.  His results can be derived from ours, although his proof is simpler.

Any length-$\ell$ path can be viewed as a subgraph of the hypercube $\{0,1\}^\ell$ by mapping
its vertices to the points $0^i1^{\ell-i}$, $0\le i\le\ell$ (here superscripting stands for repetition of coordinates).  Similarly, finite portions $\{0,1,\ldots\ell\}^d$ of the integer lattice can be mapped isometrically to a hypercube $\{0,1\}^{d\ell}$ by applying the above $0^i1^{\ell-i}$ embedding separately to each lattice coordinate.  Since isometric embedding is transitive, the graphs with finite lattice dimension are exactly the isometric hypercube subgraphs, also known as {\em partial cubes}.
The partial cube representation of a graph is unique up to cube symmetries~\cite{Lau-TR-94},
and a polynomial time algorithm for finding such representations is known from the work of Djokovic~\cite{Djo-JCTB-73,Lau-TR-94}.
Partial cubes arise naturally as the state transition graphs of {\em media}, systems of states
and state transitions studied by Falmagne et al.~\cite{cs.DS/0206033,FalOvc-DAM-02} that arise in political choice theory and that can also be used to represent many familiar geometric and combinatorial systems such as hyperplane arrangements.

The integer lattice can be viewed as a Cartesian product of paths; instead, one could consider products of other graphs.  Thus, for instance, one could similarly define the {\em tree dimension} of a graph to be the minimum $k$ such that the graph has an isometric embedding into a product of $k$ trees.
The graphs with finite tree dimension are again just the partial cubes.
Chepoi et al.~\cite{CheDraVax-SODA-02} showed that certain graph families have bounded tree dimension, and used the corresponding product representations as a data structure to answer distance queries in these graphs.    Recognizing graphs with tree dimension $\le k$ is polynomial for $k=2$~\cite{Dre-AM-92}, but NP-complete for any $k>2$~\cite{BanVel-PLMS-89}.

\section{The Semicube Graph}

\begin{figure}[t]
\centering
\includegraphics[width=5in]{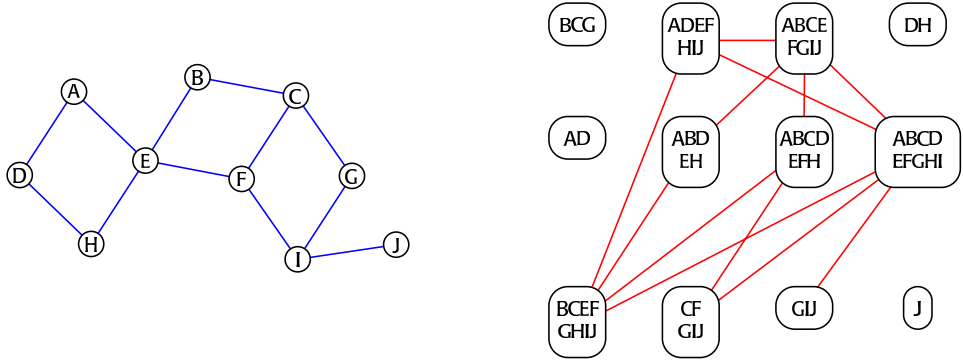}
\caption{A partial cube graph (left) and its semicube graph (right, with four isolated vertices and a connected component of eight vertices).  The six-dimensional  hypercube embedding of the graph is not shown.}
\label{fig:semicube}
\end{figure}

Throughout this paper when discussing hypercubes $\{0,1\}^\tau$ and integer lattices
$\Z^d$, we always use the $L_1$ metric, in which the distance between two points is the sum of absolute values of differences of their coordinates.

As discussed above, any graph with finite lattice dimension is a partial cube,
and polynomial time algorithms are already known for constructing partial cube representations of graphs.  Therefore, we can assume without loss of generality that we are given both an undirected graph $G$ and an isometry $\mu:G\mapsto\{0,1\}^\tau$ from $G$ to the hypercube $\{0,1\}^\tau$ of dimension $\tau$.
Let $\mu_i:G\mapsto \{0,1\}$ map each vertex $v$ of $G$ to the $i$th coordinate of $\mu(v)$.
We assume without loss of generality that $\mu$ is {\em full-dimensional}; that is, that each coordinate $\mu_i$ takes on both value $0$ and $1$ for at least one point each; for, if not, we could safely drop some of the coordinates of $\mu$
and produce a lower-dimensional hypercube isometry.
If $\mu:G\mapsto\{0,1\}^\tau$ is full-dimensional, the parameter $\tau$ is known as the {\em isometric dimension} of~$G$.

From $G$ and $\mu$ we can define $2\tau$ distinct {\em semicubes}
$S_{i,\chi}=\{v\in V(G)\mid \mu_i(v)=\chi\}$, for any pair $i,\chi$ with
 $0\le i<\tau$ and $\chi\in\{0,1\}$.
 Note that, although defined here geometrically, these sets are the same
 as the sets $W_{u,v}$ central to Djokovic's theory, which are defined
 graph-theoretically as the sets of vertices nearer to $u$ than to $v$
 for some edge $uv$.

We now construct a new graph  $\Sc(G)$, which we call the {\em semicube graph} of $G$.
We include in $\Sc(G)$ a set of $2\tau$ vertices $u_{i,\chi}$, $0\le i<\tau$ and $\chi\in\{0,1\}$.
We include an edge in $\Sc(G)$ between $u_{a,b}$ and $u_{c,d}$
whenever $S_{a,b}\cup S_{c,d}=V(G)$ and $S_{a,b}\cap S_{c,d}\ne\emptyset$; that is, whenever the corresponding two
semicubes cover all the vertices of $G$ non-disjointly.
An example of a partial cube $G$ and its semicube graph $\Sc(G)$ is shown in Figure~\ref{fig:semicube}.

As discussed earlier, a full-dimensional isometry from $G$ to a hypercube is unique up to symmetries of the hypercube.  Further, any such symmetry acts on the family of semicubes by permuting them, so the semicube graph is uniquely defined up to graph isomorphism by the graph $G$ itself, without reference to a specific isometry~$\mu$.

\section{From Lattice Embeddings to Matchings}

Suppose we are given a graph $G$ and an isometry $\lambda:G\mapsto\Z^d$ from $G$ to an integer lattice.  We use the standard Djokovic technique to embed (a finite subset of) the lattice, and therefore $G$, into a hypercube.  However we elaborate the details here and in the next two lemmas as we need the notation.
Let $\lambda_i(v)$ denote the $i$th coordinate of $\lambda(v)$,
let $\alpha_i=\min\{\lambda_i(v)\mid v\in G\}$,
let $\beta_i=\max\{\lambda_i(v)\mid v\in G\}$,
and let $\tau=\sum_i (\beta_i-\alpha_i)$.
From $\lambda$ we construct an isometry $\mu:G\mapsto \{0,1\}^\tau$
from $G$ to a hypercube, using the following construction:
for each pair of integers $i,\gamma$ satisfying
$\alpha_i\le \gamma<\beta_i$,
let $j_{i,\gamma}=\gamma-\alpha_i+\sum_{k<i}(\beta_k-\alpha_k)$;
then $j_{i,\gamma}$ uniquely identifies the pair $i,\gamma$.
We set the $j$th coordinate $\mu_j$ of the hypercube isometry to be
$\mu_j(v)=0$ if $\lambda_i(v)\le\gamma$, and $\mu_j(v)=1$ otherwise.
The map $\mu$ is then formed by using these functions as coordinates:
$\mu(v)=(\mu_0(v),\mu_1(v),\ldots,\mu_{\tau-1}(v))$.

\begin{lemma}\label{lem:m-iso}
The map $\mu$ defined as above is a full-dimensional isometry from $G$ to a hypercube.
\end{lemma}

\begin{proof}
For each coordinate $j=j_{i,\gamma}$,
$\mu_j(v)=0$ whenever $\lambda_i(v)=\alpha_i$,
and $\mu_j(v)=1$ whenever $\lambda_i(v)=\beta_i$,
so $\mu$ is full-dimensional.

If $u$ and $v$ are two vertices of $G$,
with $\lambda_i(u)<\lambda_i(v)$,
then $\mu(u)$ and $\mu(v)$ differ in the positions
$\mu_j$ where $j=j_{i,\gamma}$, $\lambda_i(u)\le\gamma<\lambda_i(v)$,
and conversely.  Therefore, the sum of the absolute values of the differences of coordinates
$\lambda_i$ is equal to the number of differing coordinates $\mu_j$,
and since $\lambda$ is an isometry, $\mu$ must also be an isometry.
\end{proof}

For any pair $i,\gamma$ with $\alpha_i\le\gamma<\beta_i$,
let $L_{i,\gamma}=\{v\mid\lambda_i(v)\le\gamma\}$ and
let $U_{i,\gamma}=\{v\mid\lambda_i(v)>\gamma\}$.

\begin{lemma}
\label{lem:lattice-semicubes}
The sets $L_{i,\gamma}$ and $U_{i,\gamma}$ described above
are semicubes of the graph $G$, and all $G$'s semicubes are of this form.
\end{lemma}

\begin{proof}
Due to the uniqueness of full-dimensional hypercube isometries,
the semicubes of $G$ are exactly those of the hypercube isometry $\mu$ constructed above.
For $j=j_{i,\gamma}$, we have that
$S_{j,0}=L_{i,\gamma}$ and $S_{j,1}=U_{i,\gamma}$.
Therefore, each $L_{i,\gamma}$ and $U_{i,\gamma}$ is a semicube,
and each semicube $S_{j,\chi}$ is of this form.
\end{proof}

It is also trivial to verify the correctness of Lemma~\ref{lem:lattice-semicubes} using
Djokovic's definition $W_{a,b}$ in place of the geometric definition of semicubes.

A {\em matching} in a graph is a collection of edges such that each vertex in the graph is incident to at most one edge of the collection.  If $M$ is a matching, we let $|M|$ denote the number of edges in $M$.

\begin{lemma}
\label{lem:match-from-iso}
If $G$ is a graph with an isometry $\lambda:G\mapsto\Z^d$,
and $\lambda_i$, $\alpha_i$, $\beta_i$, and $\tau=\sum_i(\beta_i-\alpha_i)$ are as defined above,
then there exists a matching $M$ in the semicube graph $\Sc(G)$,
such that $d=\tau-|M|$.
\end{lemma}

\begin{proof}
For every $i,\gamma$ with $\alpha_i<\gamma<\beta_i$,
we include in $M$ an edge from $U_{i,\gamma-1}$ to
$L_{i,\gamma}$.  These two semicubes together cover all of $G$,
and their intersection is the set of vertices $v$
for which $\lambda_i(v)=\gamma$; this set is nonempty because every partial cube
must be connected.  Therefore, $M$ is indeed an edge of $\Sc(G)$,
and clearly, each semicube of $G$ is associated with at most one edge of $M$.

For each coordinate $i$, $M$ includes $\beta_i-\alpha_i-1$ edges,
so the total number of edges in $M$ is
$|M|=\sum_i (\beta_i-\alpha_i-1)=\tau-d$, as was claimed.
\end{proof}

\section{From Matchings to Lattice Embeddings}

\begin{figure}[t]
\centering
\includegraphics[width=5in]{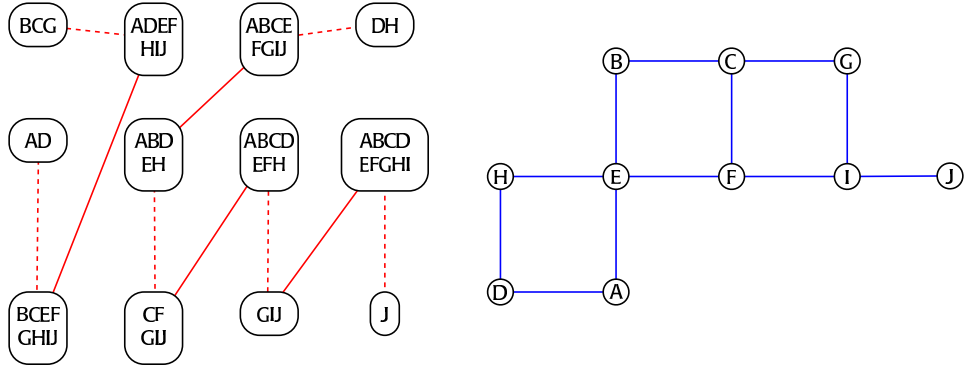}
\caption{A matching in the semicube graph (left, solid edges) completed to a set of paths by adding edges from each semicube to its complement (left, dashed edges), and the corresponding lattice embedding of the original graph (right).}
\label{fig:matched}
\end{figure}

Suppose we are given a partial cube $G$ and a matching $M$ in $\Sc(G)$.
Each vertex in $\Sc(G)$ corresponds to a semicube; we
augment $M$ to a graph $P$ (not a subgraph of $\Sc(G)$ by
adding an edge between each pair $(u,v)$ such that $u$ and $v$
correspond to complementary semicubes.
Figure~\ref{fig:matched} (left) depicts a matching $M$ and augmented graph $P$
for the semicube graph shown in Figure~\ref{fig:semicube}.
In $P$, each vertex is incident either to one edge (connecting it to its complement) or to two edges (connecting it to its complement and its match).

\begin{lemma}
\label{lem:path-superset}
If one starts from a vertex $u$ of $P$, follows an edge in $P$ to its complement $v$, and then follows another edge in $P$ to a vertex $w$ where $v$ and $w$ are matched in $M$,
then $w$ must correspond to a superset of the set corresponding to $u$.
\end{lemma}

\begin{proof}
This follows from the definition of which pairs of vertices are connected by edges in $\Sc(G)$,
and the fact that each edge in $M$ must belong to $\Sc(G)$.
\end{proof}

$V(G)$ is finite, so its subsets have no infinite ascending chain.
Any path in $P$ consists of edges that alternate in the pattern described in Lemma~\ref{lem:path-superset}, so one cannot keep following such chains of vertices indefinitely and $P$ has no cycles.
Since $P$ is a graph with no cycles, in which every vertex has degree one or two,
$P$ must consist of a disjoint union of paths $P_i$, $0\le i<d$ for some $d$.
Each path $P_i$ has an odd number of edges, since it starts and ends with an edge connecting a vertex to its complement.

Choose arbitrarily an orientation for each path, and number the vertices of $\Sc(G)$ so that
$v_{i,j}$ denotes the $j$th vertex of path $P_i$.  We let $S_{i,j}$ denote the semicube corresponding to $v_{i,j}$, and let $\ell_i$ denote the number of edges in path $P_i$.  For completeness,
let $S_{i,-1}=S_{i,\ell_i+1}=V(G)$; these subsets are not semicubes.

\begin{lemma}
For each vertex $v\in V(G)$, and each $i$, there is a unique value $x$
with $0\le x\le\lceil\ell_i/2\rceil$ and $v\in S_{i,2x-1}\cap S_{i,2x}$.
\end{lemma}

\begin{proof}
If $v\in S_{i,0}$, we are done, with $x=0$: $v$ belongs to $S_{i,-1}\cap S_{i,0}=S_{i,0}$,
and (by Lemma~\ref{lem:path-superset}) for each $j>0$, $v$ belongs to $S_{i,2j-2}$ and therefore does not belong to the complementary set $S_{i,2j-1}$.

Next, suppose that $v\notin S_{i,0}$ but $v\in S_{i,2m}$ for some integer $x>0$; let $x$ be the smallest index for which this is true. Then,
because $v\notin S_{i,2x-2}$, $v$ must belong to the complementary set
$S_{i,2x-1}$ so $v\in S_{i,2x-1}\cap S_{i,2m}$.  The same application of
Lemma~\ref{lem:path-superset} as above shows that $v$ does not belong to
$S_{i,2j-1}$ for any $j>x$.

Finally, if $v$ does not belong to any $S_{i,2j}$, then in particular it does not belong to
$S_{i,\ell_i-1}$, so it does belong to the complementary set $S_{i,\ell_i}$ and the result holds with $x=\lceil\ell_i/2\rceil$.
\end{proof}

Let $\lambda_i(v)$ denote the value $x$ found by the lemma above
for vertex $v$ and path $i$.

\begin{lemma}\label{lem:lattice-from-match}
Suppose we are given a partial cube $G$ and a matching $M$ in $\Sc(G)$.
Let $\tau$ be the dimension of any full-dimensional isometry of $G$ to a hypercube.
Then there is an isometry $\lambda:G\mapsto\Z^d$ from $G$ to an integer lattice,
with $d=\tau-|M|$.
\end{lemma}

\begin{proof}
There are $2\tau$ semicubes of $G$, of which $2|M|$ are matched in $M$.
There are two endpoints per path in $P$, which must consist of all the remaining
$2\tau-2|M|$ unmatched vertices in $\Sc(G)$.  Therefore, the number of paths in $P$ is
$d=\tau-|M|$, and the function
$\lambda(v)=(\lambda_0(v),\lambda_1(v),\ldots\lambda_{d-1}(v))$
maps $G$ to $\Z^d$.  It remains to verify that this function is an isometry.

Suppose that, for two vertices $u$ and $v$ and index $i$,
$\lambda_i(v)-\lambda_i(u)=k_i>0$.
Then, among the semicubes corresponding to vertices on path $P_i$,
the ones containing $u$ but not containing $v$ are
$S_{i,2\lambda_i(u)}$, $S_{i,2\lambda_i(u)+2}$, $\ldots$, $S_{i,2\lambda_i(v)-2}$;
there are exactly $k_i$ such semicubes.  By a symmetric argument we can find
$k_i$ semicubes containing $u$ but not $v$ when
$\lambda_i(v)-\lambda_i(u)=-k_i<0$.
Summing over all choices of $i$, this means that there are exactly $k$ semicubes of $G$
that contain $u$ but do not contain $v$, where $k=\sum_i k_i$ is the $L_1$ distance between $\lambda(u)$ and $\lambda(v)$.  However, it follows from the definition of $L_1$ distance in a hypercube that, for any vertices $u$ and $v$ in a partial cube, the distance between $u$ and $v$ equals the number of semicubes that contain $u$ but do not contain $v$.
Therefore, the distance between $u$ and $v$ in $G$ equals their distance in $\lambda(G)$ and $\lambda$ is an isometry.
\end{proof}

A two-dimensional lattice embedding for the partial cube of Figure~\ref{fig:semicube}
is shown in Figure~\ref{fig:matched} (right).

\section{The Main Result}

\begin{figure}[t]
\centering
\includegraphics[width=5in]{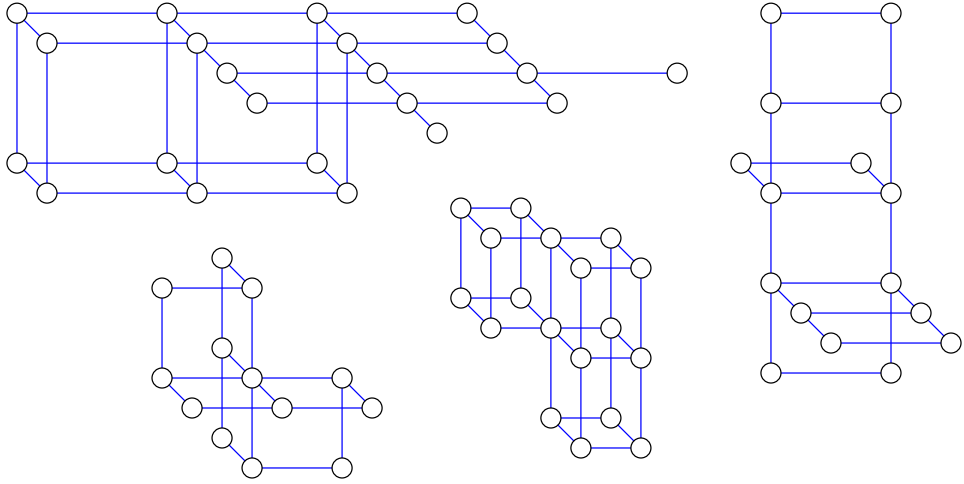}
\caption{Four examples of three-dimensional lattice embeddings found by our implementation of our lattice dimension algorithm.}
\label{fig:grids3d}
\end{figure}

\begin{theorem}
If $G$ is a partial cube with isometric dimension $\tau$, then the lattice dimension of $G$ is
$d=\tau-|M|$ where $M$ is any maximum matching in $\Sc(G)$.
\end{theorem}

\begin{proof}
This follows immediately from Lemmas \ref{lem:match-from-iso} and~\ref{lem:lattice-from-match}.
\end{proof}

In our algorithm analysis, as in~\cite{cs.DS/0206033},
we use $n$ to stand for the number of vertices of an input graph $G$,
$m$ to stand for its number of edges, and $\tau$ to stand for the isometric dimension of~$G$.
As in~\cite{cs.DS/0206033}, we may use the inequalities
$m\le n\log_2 n$ and $\log_2 n\le\tau<n$ to aid in the comparison of time bounds involving these quantities.

\begin{theorem}
If we are given a partial cube $G$, and a full-dimensional hypercube isometry
$\mu:G\mapsto\{0,1\}^\tau$, we can compute in time $O(n\tau^2)$ the lattice dimension $d$ of $G$,
and in the same time construct a lattice isometry
$\lambda:G\mapsto\Z^d$.
If we are given only $G$, and not its hypercube isometry, we can perform the same
tasks in time $O(mn+n\tau^2)$.
\end{theorem}

\begin{proof}
We construct the semicube graph $\Sc(G)$ directly from the definition, by testing each pair of semicubes, in time $O(n\tau^2)$, and use a maximum matching algorithm to find a matching with the largest possible number of edges in $\Sc(G)$, which can be done in time
$O(\tau^{2.5})$~\cite{MicVaz-FOCS-80}.
It is then straightforward to apply the construction of Lemma~\ref{lem:lattice-from-match}
to transform the matching into a lattice isometry with dimension $d=\tau-|M|$,
in time $O(n\tau)$.
The total time is dominated by the $O(n\tau^2)$ bound for finding $\Sc(G)$.
If we are not given $\mu$, we can construct it using the method of Djokovic in time $O(mn)$~\cite{Djo-JCTB-73,Lau-TR-94}.
\end{proof}

\section{Conclusions}

We have described a polynomial time algorithm for finding the minimum lattice dimension of a graph.

We implemented this algorithm as part of a system for visualizing media, using the Python programming language,
however for finding maximum matchings our implementation replaces the somewhat complex matching algorithm of
Micali and Vazirani~\cite{MicVaz-FOCS-80} with the slower but somewhat less complex blossom-contraction algorithm of Edmonds~\cite{Edm-CJM-65}.
Our implementation takes as input a description of a medium,
and produces as output a drawing of the corresponding partial cube,
embedded into a lattice of minimum dimension;
some examples of its output are shown in Figure~\ref{fig:grids3d}.

As well as its applications in graph visualization, we believe that our algorithm may be useful in constructing concise labeling schemes for partial cubes that enable fast distance and routing queries.

We note that finer control over the lattice embedding produced by our algorithm may be available, by removing some edges of $\Sc(G)$ before applying a matching algorithm, or by giving the edges weights representing the desirability of making certain dimensions line up and by using a weighted maximum matching algorithm.  For instance, the former approach can be used if we are searching for a lattice embedding of an oriented graph in which the embedding must assign the tail of each oriented edge a lower coordinate value than its head.  However, more remains to be done on finding ways to choose among multiple matchings in $\Sc(G)$ and the corresponding multiple possible lattice embeddings of a graph, to select the one most suitable for a given application.  For instance, when drawing a partial cube,
it may be of interest to choose a lattice embedding that maximizes the amount of symmetry of the drawing, and additional work would be needed to incorporate such symmetry display considerations into our matching algorithm.

It would also be of interest to find more efficient algorithms for constructing the semicube graph, as that is the major time bottleneck of our algorithm, and to investigate more carefully the combinatorial properties of this graph.

\section*{Acknowledgements}

This research was supported in part by NSF grant CCR-9912338.
I thank Jean-Claude Falmagne for discussions that led to these results, and Sergei Ovchinnikov for encouraging me to publish these results and for many helpful suggestions
on drafts of this paper.

\raggedright
\bibliographystyle{abuser}
\bibliography{media}

\end{document}